# Petrographic and geochemical evidence for multiphase formation of carbonates in the Martian orthopyroxenite Allan Hills 84001

Carles E. MOYANO-CAMBERO[a,*], Josep M. TRIGO-RODRÍGUEZ[a], M. Isabel BENITO[b], Jacinto ALONSO-AZCÁRATE[c], Martin R. LEE[d], Narcís MESTRES[e], Marina MARTÍNEZ-JIMÉNEZ[a], F. Javier MARTÍN-TORRES[f,g], and Jordi FRAXEDAS[h]

[a] Institute of Space Sciences (IEEC-CSIC), Campus UAB, Carrer de Can Magrans, s/n, 08193 Cerdanyola del Vallès (Barcelona), Spain (moyano@ice.csic.es, trigo@ice.csic.es, mmartinez@ice.csic.es).

[b] Departamento de Estratigrafía-IGEO, Facultad de Ciencias Geológicas, Universidad Complutense de Madrid-CSIC, José Antonio Nováis, 12, 28040 Madrid, Spain (mibenito@ucm.es).

[c] Fac. de Ciencias Ambientales y Bioquímica. Avda. Carlos III s/n. Universidad de Castilla-La Mancha. . 45071 Toledo, Spain (jacinto.alonso@uclm.es).

[d] School of Geographical and Earth Sciences, University of Glasgow, Gregory Building, Lilybank Gardens, Glasgow G12 800 (martin.lee@glasgow.ac.uk).

[e] Institut de Ciència de Materials de Barcelona (ICMAB-CSIC), Campus UAB, 08193 Bellaterra (Barcelona), Spain (narcis@icmab.es).

[f] Instituto Andaluz de Ciencias de la Tierra (CSIC-UGR), 18100 Armilla, Granada, Spain, (javiermt@iact.ugr-csic.es).

[g] Division of Space Technology, Department of Computer Science, Electrical and Space Engineering, Luleå University of Technology, Kiruna, Sweden.

[h] Catalan Institute of Nanoscience and Nanotechnology (ICN2), CSIC and the Barcelona Institute of Science and Technology, Campus UAB, 08193 Bellaterra (Barcelona), Spain (jordi.fraxedas@icn2.cat).

[*] Corresponding author.

**Abstract** – Martian meteorites can provide valuable information about past environmental conditions on Mars. Allan Hills 84001 formed more than 4 Gyr ago, and owing to its age and long exposure to the Martian environment, this meteorite has features that may record early processes. These features include a highly fractured texture, gases trapped during one or more impact events or during formation of the rock, and spherical Fe-Mg-Ca carbonates. Here we have concentrated on providing new insights into the context of these carbonates using a range of techniques to explore whether they record multiple precipitation and shock events. The petrographic features and compositional properties of these carbonates indicate that at least two pulses of Mg- and Fe-rich solutions saturated the rock. Those two generations of carbonates can be distinguished by a very sharp change in compositions, from being rich in Mg and poor in Fe and Mn, to being poor in Mg and rich in Fe and Mn. Between these two generations of carbonate is evidence for fracturing and local corrosion.

**INTRODUCTION**

Most Martian meteorites are igneous achondrites. A possible exception is Northwest Africa (NWA) 7034, whose classification is debated, although it is officially described as a basaltic breccia (Agee et al., 2013). Historically the Martian meteorites have been referred to as 'SNC', although it is now known that some of them do not fit into the three traditional families: shergottites, nakhlites, chassignites (Mittlefehldt, 1994; Agee et al., 2013). Martian meteorites have a characteristic mineralogy, bulk chemistry, and oxygen isotope composition (with NWA 7034 and its pairs being an exception, according to Agee et al., 2013) that distinguish them from other extraterrestrial rocks (McSween, 1994; Treiman et al., 2000). The incontrovertible evidence of their Martian origin is the isotopic composition of trapped gas in shock-formed glass, which matches that of the planet's atmosphere as measured by the Viking lander (Owen, 1977; Bogard and Johnson, 1983).

Here we have studied Allan Hills (ALH) 84001, a 1.93 kg meteorite that was found in Antarctica in 1984. It was ejected from Mars 15 million years ago (Nyquist et al., 2001), and fell to Earth ~13,000 years ago (Jull et al., 1995). Several different crystallization ages have been reported for ALH 84001 (Nyquist et al., 2001; Lapen et al., 2010), thus highlighting the difficulty in dating a meteorite in an open system with a complex post-crystallization history including aqueous alteration and shock metamorphism (Treiman, 1998). It is accepted by the scientific community that ALH 84001 is the oldest known Martian meteorite. It formed ~4.1 Gyr ago (calculated using $^{176}$Lu-$^{176}$Hf chronometry by Lapen et al., 2010), and during a period of intense bombardment (Frey, 2008). Therefore, ALH 84001 is a fragment of Noachian crust, which forms the ancient highlands of Mars (Hamilton et al., 2003). As this meteorite is the only sample we have of the oldest Martian crust, it is a very valuable source of information about the earliest conditions and subsequent evolution of Mars (Bridges et al., 2001).

Once it was recognized as a Martian meteorite, ALH 84001 was classified as an orthopyroxenite (the only one found so far), and therefore is not part of the traditional SNC family (Mittlefehldt, 1994). It contains 97 vol. % of orthopyroxene ($Fs_{27.3}Wo_{3.3}En_{69.4}$) as grains up to 6 mm in size. Other constituents include 2 vol. % of euhedral to subhedral chromite clasts, and 1 vol. % of augite, olivine, pyrite, apatite, plagioclase (mostly transformed into maskelynite or glass, by shock), phosphate and $SiO_2$ (Mittlefehldt, 1994). The minor minerals are mostly remnants of trapped interstitial melt.

ALH 84001 has a highly fractured texture (Fig. 1), most probably due to progressive deformation by impacts while it was still on Mars (Treiman, 1995). In fact, the meteorite has a complicated shock history, which may have moved it progressively closer to the spallation zone of the impact that ejected it (Melosh, 1984). At least two impact events are responsible for its shock features (Treiman, 1995); the first event led to brecciation and relocation of the rock closer to the planet's surface, and the second higher intensity event transformed plagioclase into maskelynite or shock-modified plagioclase glass. As a consequence, it has been affected by shock-induced metamorphism. The orthopyroxene and chromite in ALH 84001 are in cumulus phases due to the formation of discrete crystals forming a primary precipitate during slow cooling in a surrounding magma, conforming thus to an orthopyroxene adcumulate (for a description of igneous cumulates see e.g. Hall, 1987). The uniform chemical composition of the pyroxene could indicate that ALH 84001 cooled more slowly than other Martian meteorites during magmatic or metamorphic recrystallization (Mittlefehldt, 1994), probably at a depth of ~40 km (Kring and Gleason, 1997).

One of the most interesting features of ALH 84001 is the presence of Mg-Fe-Ca carbonates along fractures and in cataclastic areas (Scott et al., 1998). The study of Martian carbonates is important as these minerals precipitate from aqueous fluids and thus can be used to explore the history of water at the planet's surface and crust (Bridges et al., 2001; Tomkinson et al., 2013). Carbonate outcrops have been found on the Martian surface by orbiters, landers and rovers (Bandfield et al., 2003; Ehlmann et al., 2008; Boynton et al., 2009; Morris et al., 2010). Carbonate deposits that may represent sites of paleolakes support the idea that Mars had a warmer and wetter past (Wray et al., 2011). In particular, the amount of carbonates present at the surface are a powerful tool for understanding the duration of the warmer and wetter period of Martian history. Additionally, carbonates can contain microstructures that form by high shock pressures and temperatures, thus enabling them to be used to explore the magnitude and frequency of impact events (Treiman, 1998).

The carbonates in ALH 84001 are known to be Martian because they are cut by and predate some of the fractures mentioned previously, and so have received considerable attention. The detailed study by Corrigan and Harvey (2004) described them as "rosettes" (here termed "globules"), planiform "slabs", "post-slab" magnesites, and as interstitial carbonates, but they have been classified in several other ways (e.g. Saxton et al., 1998; Scott et al., 1998; Holland et al., 1999; Corrigan et al., 2003; Niles et al., 2005; Meyer, 2012, and references therein). One early interpretation of their origin was that it was associated with early biologic activity on Mars (McKay et al., 1996), but subsequent studies have shown that their precipitation can also be explained by abiotic processes (Scott et al., 1998; Warren, 1998; Golden et al., 2001; Treiman et al., 2002; Corrigan and Harvey, 2004; Niles et al., 2006; 2009; Steele et al., 2007; Martel et al., 2012; Meyer, 2012, and references therein). However, what make these carbonates objectively interesting are their formation process and their ~3.9 Gyr age (calculated using $^{87}$Rb-$^{86}$Sr chronometry by Borg et al., 1999). The anomalies in magnetization of these carbonates could indicate that the Martian magnetic field shut off between the formation of these carbonates and the transition from the late Noachian to Hesperian (Weiss et al., 2002). ALH 84001 is not the only Martian meteorite containing carbonates, since they have also been described from the nakhlites (e.g., Carr et al., 1985; Chatzitheodoridis and Turner, 1990; Bridges and Grady, 2000; Bridges and Schwenzer, 2012; Tomkinson et al., 2013). However, and although the formation process for those carbonate could be fairly similar (Bridges and Grady, 2000; Bridges et al., 2001), their petrography and composition are clearly different (Lee et al., 2013; Tomkinson et al., 2013). This is why we focus here in the study of the carbonates in ALH 84001, since they are much older and, therefore, provide information about the conditions on Mars long before the carbonates found on nakhlites were formed (Bridges et al., 2001).

The distribution of carbonates throughout ALH 84001 provides clues about the environmental conditions under which they formed, and can help to establish a link to the carbonates that have been found at the surface of Mars. Following several studies, two main formation scenarios at ~3.9 Gyr have been proposed: (i) precipitation at low temperature (e.g. Treiman, 1995; Romanek et al., 1995; McKay et al., 1996; Valley et al., 1997; Golden et al., 2000; Eiler et al., 2002; Theis et al., 2008; Halevy et al., 2011); (ii) precipitation from a fluid with high concentrations of $CO_2$, possibly over a short time, at high temperatures (e.g. Mittlefehldt, 1994; Harvey and McSween, 1996; Scott et al., 1997a; 1998). This second scenario may have been the consequence of an impact event and subsequent hydrothermal activity (Harvey and McSween, 1996; Treiman et al., 2002), or crystallization of shock melted material (Scott et al., 1997a; 1998). The terrestrial carbonates described in Treiman et al. (2002) support, by analogy, formation due to Martian hydrothermal processes. Indeed, the carbonates found in ALH 84001 are compositionally different from the common carbonates (or any of their solid solutions) found on Earth. In fact, there are just a few regions on Earth where similar (but not identical) carbonates occur (Treiman et al., 2002), and they are out of the range of terrestrial carbonate solids solutions. Also, our studies provide complementary information for the multi-generational carbonate model that was described by previous authors (e.g. Corrigan and Harvey, 2004). Our results provide textural and compositional evidence for precipitation of the carbonates by at least two different fluids.

**EXPERIMENTAL TECHNIQUES**

ALH 84001 was found in the Far Western Icefield of Allan Hills by the Antarctic Search for Meteorites (ANSMET) program in 1984. A preliminary study provided a weathering degree of A/B and a 'fracturing category' of B (equivalent to an S4-5 shock grade of the classification used currently)(Score and MacPherson, 1985). The NASA Johnson Space Center (JSC), responsible for the curation of the ANSMET collection, provided the 30 µm thin section ALH 84001,82 (see Meyer, 2012, for the specific location of this sample in the original rock). In order to comprehensively characterise carbonates in this sample we have used transmitted light (TL) and cathodoluminescence (CL) microscopy, electron microprobe analysis, micro-Raman spectroscopy, scanning electron microscopy (SEM) and Energy-Dispersive X-ray Spectroscopy (EDS). The reasons why we have chosen each technique and its particular relevance to this study are explained below.

**Petrographic microscopy**

We used three petrographic microscopes. At the Institute of Space Sciences we worked with a Zeiss Scope, with magnifications up to 500×, and with a Motic BA310Pol Binocular microscope, which has a rotating platen that enables viewing of features including undulatory extinction. Both can work with crossed nicols. The third microscope, at the Universidad Complutense de Madrid, is a Nikon Eclipse LV100NPol, working with magnifications up to 600×. Summing the capabilities of the three of them we are able to work with reflected light (RL), transmitted light (TL, particularly useful to recognize features on samples with a size of tens of µm), TL under crossed nicols (for identifying minerals), and with bright field (high contrast and differentiation between more and less dense materials), or dark field illumination (use of scattered light to increase the visibility of low relief features). These techniques enable a great deal of information to be obtained from the regions of interest. We used these microscopes to take 500×images of the ALH 84001,82 section. In order to have a reference map we merged the TL images in a high-resolution mosaic (Fig. 1). A grid was superimposed to the image (square size is 1 mm$^2$) in order to easily navigate around the section and so (re)locate regions of interest.

**Cathodoluminescence microscopy**

With regards to CL, the light emitted during electron bombardment of a mineral has an intensity and wavelength that depends on its chemical composition and crystallographic structure (Burns, 1969). Even trace amounts of some elements can act as activators or quenchers of CL (Götze et al., 2000, Akridge et al., 2004). The energy of the electron beam can also have an influence on CL emission (Akridge et al., 2004). Therefore to correctly interpret CL results it is important to constrain the mineralogy of the sample being studied. Iron-free minerals can be particularly luminescent, as iron is a typical quencher (Burns, 1969). With this property in mind, carbonates can be identified in the thin section as they are zoned, with layers that are Mg-rich and Fe-free (Saxton et al., 1998), which should luminesce red (Akridge et al., 2004). In this study, we used a Technosyn cold cathodoluminescent MK4 operated at 20–24 kV/350–400 mA, at the Universidad Complutense de Madrid. The CL was used to create a mosaic of the thin section together with higher magnification and resolution images (Figs. S1 and 2).

**Scanning Electron Microscopy**

Two SEMs were used: a FEI Quanta 650 FEG with a Back Scattered Electron Detector (BSED), at the Institut Català de Nanociència i Nanotecnologia (ICN2), and a JEOL JSM7600F at the National Center for Electron Microscopy (ICTS) at the Universidad Complutense de Madrid. The Quanta 650 was operated in low-vacuum mode with the thin section uncoated so that any carbon detected by EDS analysis was not from the coating. Secondary Electron (SE) and Back-Scattered Electron (BSE) imaging was used to study in detail sites of interest in the sample, and at magnifications of up to 27,000× (see Figs. 3 to 6). The SEM was also used to select regions and points of interest for subsequent chemical analysis by electron microprobe.

**Energy-Dispersive X-ray Spectroscopy**

EDS was used to create elemental maps of the regions of interest (Fig. 3). The EDS detector was an Inca 250 SSD XMax20 with a detector area of 20 mm$^2$, attached to the SEM at the Institut Català de Nanociència i Nanotecnologia (ICN2). Since we had available the electron microprobe technique, EDS was mostly used to create the illustrative chemical maps (Fig. 3), instead of performing quantitative chemical analyses. The accelerating voltages were 20 kV and the totals were normalized to 100%.

**Micro-Raman spectroscopy**

Micro-Raman spectra were acquired from the carbonates using a Jobin-Yvon T-64000 Raman spectrometer at the Institut de Ciència de Materials de Barcelona (ICMAB-CSIC). The instrument was attached to an Olympus microscope and equipped with a liquid nitrogen-cooled CCD detector, and was operated in backscattering geometry at room temperature using the 5145 Å line of an argon-ion laser. The lateral spatial resolution was ~ 1 μm, and the laser power onto the sample was maintained between 0.6 and 0.7 mW to avoid degradation or excessive heating. High-resolution spectra were acquired in working windows between 100 and 1,400 cm$^{-1}$. The instrument was used to better characterize the composition of the carbonates and possibly also to detect shock features (Fig. 4).

**Electron microprobe**

Quantitative chemical analyses and BSE images were obtained using a JEOL JXA-8900 electron microprobe equipped with five wavelength-dispersive spectrometers at the Universidad Complutense de Madrid.

More than 160 spot analyses were acquired from the carbonate globules (Fig. 5) using an accelerating voltage of 15 kV, beam current of 10 nA, and a spot size of 1 μm; line profiles used 15 kV/100 nA, and spot size and step interval of 1 μm diameter (dwell time =1000 msec). The counting time for spot analyses was 30 seconds per element, although 5 elements could be analysed simultaneously. The relatively low voltage, beam current and counting time employed for the carbonate spot analyses minimized beam damage. The standards used were siderite for Fe and Mn, and dolomite for Mg and Ca. Measured precision was better than ± 0.0014% for Mg, ± 0.009% for Ca, ± 0.005% for Fe, and ± 0.19% for Mn. Detection limits were 150–180 ppm for Mg, 150–200 ppm for Ca, 350–500 ppm for Fe and 400–510 ppm for Mn. Elemental compositions were given as element oxides % and were recalculated to mol% of $MgCO_3$, $CaCO_3$, $FeCO_3$ and $MnCO_3$.

**RESULTS**

Our work has focused on carbonates in the regions marked D5 and J7 on the TL and CL high-resolution mosaics (Figs. 1, 2 and S1). D5 contains abundant carbonate globules attached to the walls of a former pore (Fig. 2), while J7 is a large globule that lies next to a granular band (Fig. 2), and has a rhombohedral core (J7.3 in Figs. 3-5 and 7). In these regions, both of which are close to fractured areas, the space between carbonate globules contains maskelynite, which has probably formed by shock transformation of plagioclase (Fig. 2)

**Petrographic context and microstructures of the carbonates**

Two main fracture orientations can be identified in Figure 1; one is aligned NNW-SSE, and the other ENE-WSW. As the ENE-WSW fractures cross-cut and displace the other set, they must be younger (Fig. 1). CL imaging highlights some of the layers in carbonates by their high intensity of red luminescence in comparison to the rest of the sample (which is largely non-luminescent; Fig. S1). The high CL intensity is inferred to be the consequence of a particularly low content of Fe, which would otherwise quench the Mn–activated luminescence (e.g. Machel et al., 1991). Areas with blue to grey CL (pale areas under TL) are maskelynite or plagioclase glass (Plg in Fig. 2). Non-luminescent grains are chromite (a few can be seen in Fig. S1). As expected, we found that carbonates are present in or close to fractures, and are also dispersed throughout the matrix-breccia (Figs. 1, 2 and S1). In regions where the two fracture systems cross-cut, globules seem to be bigger and more abundant, (e.g., D5), and plagioclase is also more abundant (Figs. 1, 2 and S1).

SEM work shows that the carbonate globules are characterized by a very clear and previously well documented zonation that divides them into several 'layers' (e.g. Saxton et al., 1998; Holland et al., 1999; Golden et al., 2001; Corrigan and Harvey, 2004). This zoning can be readily characterized by BSE imaging (Figs. 3-6) and EDS element mapping (Fig. 3). The layers are differentiated using the scheme in Fig. 7. The pattern of layering is comparable between globules, although in J7 the oldest part can be recognized (J7.3 in Figs. 3-5 and 7). A preliminary description of this 'core' was provided by Moyano-Cambero et al. (2014). It is likely to be equivalent to the central regions of other globules, but is surrounded by a thin rim that is lighter grey in BSE images (and therefore has a higher mean atomic number). The core has a rhombohedral shape, which is consistent with a carbonate mineral (i.e., calcite, siderite, magnesite, rhodocrosite, ankerite or dolomite). Hereafter we refer to the core and its rim as layers #1 and #2, respectively (see J7.3 in Fig. 7). Layer #2 is followed by a much thicker and uniform layer (#3) that is comparable to layer #1 in BSE images (see J7.6 in Figs. 6 and 7). Layer #3 is equivalent to the first-formed zone that has been distinguished in other globules found in this sample (e.g., D5.1 in Figs. 3-6), which we also call layer #3 (Fig. 7). Carbonate globules in the two selected areas (D5 and J7) gradually become darker in BSE images (lower in mean atomic number) until their euhedral terminations at the margins of layer #4 (Figs. 6, 7, ). It is interesting to note the presence of small fractures that terminate at the end of layer #4 and do not affect layer #5 (Fig. 6). Also, we observed that although the boundary between layers #4 and #5 is mostly euhedral, in some areas a small-scale corrosion can be observed affecting the outermost part of layer #4 (Fig. 6). Layer #5, distinctive in BSE images, gradually grades into layer #6, which has a lower mean atomic number. Finally, the zonation ends in a very thin and clear area (layer #7), although it is not present in all globules (Fig. 7). At the boundaries between layer #4 and #5, and between #6 and #7, are small inclusions of high atomic number, which are petrographically compatible with Fe oxides and sulfides (Fig. 6, D5.5). However, the very small size of these minerals (<500 nm) meant that we were unable to characterize them properly. Nevertheless, the areas containing these minerals have overall higher content of Fe than the surrounding carbonate, and lower concentrations of Na, Cr and locally S (less than 1%), which are not detected in the surrounding carbonate.

**Carbonate and oxide mineralogy**

The Raman spectrometer was used to analyze carbonate in J7 (Fig. 4). CrystalSleuth software (Laetsch and Downs, 2006) was used to compare the spectra obtained with those in the RRUFF catalogue (Downs, 2006), and to remove the background from our spectra. As a consequence, the particularly fluorescent spectra obtained at points H, I and J were very noisy, but interesting features could nonetheless be distinguished (Fig. 4). The main peaks obtained (Table S1) are not easily attributed to a specific carbonate mineral. As shown in Fig. 4 several minor peaks (~230, 250, 300, 415 and 1320 cm$^{-1}$) can be related to hematite, which is consistent with previous studies suggesting the presence of microcrystalline grains of this mineral within the carbonates (Steele et al., 2007). The peak that occurs in all the spectra at ~300 cm$^{-1}$, which is common in carbonates but varies with composition, can be modified by one peak of hematite existing at approximately the same wavenumber. Therefore, and although in most cases they seem similar to the dolomite peak, we cannot truly associate the exact position of this peak to that of a specific carbonate mineral (except for I and J, where the typical 330 cm$^{-1}$ peak of magnesite can be clearly recognized). The peak at ~730 cm$^{-1}$ also proved to be difficult to attribute, as most spectra show Raman shifts that are too high compared to the common values for carbonates. This is why in Fig. 4 those peaks are mainly associated with siderite, because it is the carbonate showing a highest shift for this peak. Something similar happens with the peak at ~1100 cm$^{-1}$. Apart for a minor possible influence of hematite in the peak position, all the analyzed points (apart for A, C and D) show between 2 and 4 peaks in the 1085 to 1103 cm$^{-1}$ region (Table S1). Also of interest is a peak at between 660 and 680 cm$^{-1}$ that can be clearly distinguished in several of the spectra (Fig. 4) and is interpreted to indicate the presence of magnetite, whose most prominent peak is at ~660 cm$^{-1}$. This finding of magnetite is consistent with previous studies (McKay et al., 1996; Cooney et al., 1999), although according to our Raman results it seems to be present all around the carbonate at J7 (Fig. 4), and presumably in other carbonates too.

**Carbonate compositional zoning**

EDS element maps of D5.1 (Fig. 3) show the alternating Fe-rich (red) and Mg-rich (green) nature of the layers. Electron microprobe analyses were used to quantify carbonate chemistry at ~160 points in the sample (Fig. 5) and show the progressive compositional trends in the globules. The overall composition of each layer is shown in Table S1 and the microprobe profiles and spot analyses of globules D5 and J7 are in Figs. 8 and 9, respectively. A subset of these data (the spots marked at J7.2, J7.3, D5.1 and D5.2 in Fig. 5) are listed in Table 2.

Layers #1 and #2, are visible only in the carbonate at J7 (points 3, 4 and 6 at J7.3). They differ from other zones in having intermediate concentrations of Fe and Mg, and the highest Ca compositions, plus relatively high Mn values (Figs. 8, 9 and Table 1). The main differences between the two layers are that #2 has a lower Mg, and greater Ca, Mn and Fe concentrations. The outer part of layer #2 (indicated as 'outer #2' in Fig. 7) shows high Ca and Mn compared to any other regions in the carbonates, and simultaneously very low Fe and Mg abundances (Figs. 8, 9 and Table 1, and points 27 and 5 at J7.3 in Fig. 5 and Table 2). This specific portion of layer #2 cannot be clearly resolved in BSE images, but can be easily identified as a blue ring (Ca-rich region) in the elemental map at J7.3 (Fig. 3). Layer #3 is characterized by a progressive decrease in Ca, Mn, and Fe with a concomitant enrichment in Mg (Figs. 8, 9 and Table 1; see points 7, 31, 34, 82, 84 and 118 in Fig 5 and Table 2). We have divided this compositional progression in three phases: #3.1 (the closest area to layer #2, still quite Ca- and Mn-rich), #3.2 (main layer #3, the wide area between layers #2 and #4) and #3.3 (the closest area to layer #4). Layer #4 (like in points 38, 87 and 119, Fig. 5) follows the variation in composition started in layer #3 but at greater intensity, until reaching very high contents in Mg (Figs. 8, 9). Layer #5 shows an abrupt change in globule composition compared to layer #4 (Figs. 9, 10): Mg content strongly decreases whereas Fe, Ca and Mn concomitantly increase (Figs. 9, 10; points 12, 92 and 123, Fig. 5). This sharp change in carbonate composition coincides with the presence of local corrosion and the small fractures affecting the end of layer #4 and not affecting layer #5 (Fig. 6). In layer #6 (points 49, 13, 97, 98, 126 and 128, Fig. 5) there is again a progression towards a very Mg-rich carbonate (Fig. 8, 9; points 98 and 128 in Fig. 5 and Table 1). This layer has the lowest Fe contents, thus accounting for its high luminescence intensity. Some globules have a thin Fe, Ca and Mn-rich rim (layer #7, represented by point 101, Fig. 5) which exhibits higher Fe, Ca and Mn contents than layer #6 (Figs 8, 9).

The microprobe data were plotted in a ternary diagram (Fig. 10) for comparison with the results of previous studies, with terrestrial carbonates, and with the distribution of carbonate solid solutions (Chang et al., 1996). According to the ternary diagrams (Fig. 10), the studied carbonates are not in the range of terrestrial carbonate solid solutions, and have a composition that is intermediate between dolomite, ankerite and the siderite-magnesite solid solutions. The carbonates studied here are compositionally similar to those analyzed by previous authors (Mittlefehldt, 1994; Harvey and McSween, 1996; Valley et al., 1997; Scott et al., 1998; Holland et al., 1999; 2005; Eiler et al., 2002; Corrigan and Harvey, 2004; Halevy et al., 2011), as shown in Fig. 10. However, some of those authors observed a very Ca-rich carbonate phase in bulk samples of ALH 84001, and in thin sections ALH 84001,119, ALH 84001,287, ALH 84001,302 and ALH 84001,303, which is not observed in the sample studied here, ALH 84001,82.

## DISCUSSION

### Environment of carbonate precipitation

It has been suggested that most of the ALH 84001 carbonates formed from a $CO_2$-enriched aqueous fluid penetrating an already brecciated rock (Harvey and McSween, 1996). These processes are interpreted to have coincided with impact metamorphism, perhaps with the solutions sourced from a transient impact crater lake (Warren, 1998). Formation of the carbonate in a low temperature near surface environment over short timescales has been suggested (Berk et al., 2010), maybe by a mixture of subsurface and atmosphere-derived fluids or within an alkaline spring environment (Niles et al., 2005, Halevy et al., 2011). Similar carbonates have been found only in a few places in Earth, and the presence of magnesite rather than hydromagnesite, the C and O isotope ratios consistent with near-surface reservoirs and ocean water, respectively, and the lack of uniformity in the distribution of carbonates, point toward a hydrothermal origin for them (Treiman et al., 2002). Due to the similarities between these terrestrial carbonates and those in ALH 84001, an origin in a hydrothermal system has been proposed (Treiman et al., 2002).

### Relative chronology of carbonate precipitation

There have been several interpretations of the proposed ages of these carbonates (Nyquist et al., 2001), and the present study suggests that different types, and maybe different generations, are present (Mittlefehldt, 1994). According to our petrographic and geochemical results, the formation of these carbonates can be divided in at least two phases of precipitation. The carbonates precipitated in this meteorite as zoned globules, disks, or slabs along fractures, and as interstitial material within pyroxene crystals (e.g. Corrigan et al., 2003; Corrigan and Harvey, 2004). In the studied section, carbonate globules are found mostly in regions that had been previously highly fractured (Figs. 1, 2 and S1), like the points where the two fractures systems intersect. The presence of these distinct fractures systems, with the one aligned NNW-SSE being traversed and displaced by the one aligned ENE-WSW, suggests that the sample also experienced at least two shock events. Granular bands (i.e., 'GB' in Fig. 2), have been previously interpreted as being older than the fractures (Mittlefehldt, 1997), pointing towards more than two shock events. However, we consider the granular bands to be contemporary with the fractures. Indeed, these types of bands typically form as a product of friction during terrestrial fracturing (see e.g. Brodie et al., 2007, where they are named cataclasites). These bands could actually show a high carbonate content due to their greater original porosity.

In common with previous models (e.g. Schwandt et al., 1999), we suggest that the most plausible scenario for carbonate formation is initial precipitation from Mg- and Fe-rich solutions, followed by the formation of maskelynite or shock modified plagioclase glass (Fig. 2) during a shock event (Mittlefehldt, 1994; Treiman 1995, 1998; Scott et al., 1997b; Cooney et al., 1999; Greenwood and McSween, 2001; Corrigan et al., 2003). These solutions would have filled the shock-produced fractures at different times, leading to the precipitation of carbonates in at least two events. Indeed, carbonates are partially corroded at their contact with maskelynite or plagioclase glass (Fig. 3, D5.1). The density of fractures also differs dramatically between the carbonate and glass (Mittlefehldt, 1994), which supports the suggestion that out of these two components the carbonates formed first. The presence of these solutions, along with other evidence (described, e.g., in Niles et al., 2006, and at the references therein) imply that liquid water was relatively abundant on Mars ~3.9 Gyr ago. In fact, those carbonates could be just a small sample of the products of quite a common phenomenon on the subsurface of Mars (Michalski and Niles, 2010).

The characteristic compositional zoning of ALH 840001 carbonates (e.g. Saxton et al., 1998; Corrigan and Harvey, 2004) indicates that they precipitated in a manner whereby each zone formed when saturation was reached (Halevy et al., 2011). This zonation has been described as a consequence of chemical variation of their parent fluid, driven by the carbonate precipitation since Fe-, Mn- and Ca-rich solutions are more kinetically favored to precipitate than Mg-rich solutions (Harvey and McSween, 1996; Saxton et al., 1998; Corrigan et al. 2003; Corrigan and Harvey, 2004; Niles et al., 2009, among others). The last carbonates to precipitate are therefore Mg-rich, only when the initial fluid is much more depleted in Fe, Ca and Mn than it was at the beginning. Despite other differences, this zonation is equivalent between the globules and the interstitial carbonates precipitated within pyroxene crystals (see e.g. Scott et al., 1998, for the distinction between carbonates). A similar trend in the chemical evolution of the carbonates can be observed twice (from layers #1 to #4, and from #5 to #6) (Figs. 9, 10), which together with evidence described later suggests that these carbonates precipitated in at least two events.

**Carbonate mineralogy**

None of the carbonate spectra in the RRUFF database precisely fit any of our Raman results. This finding, together with the multiplicity of peaks and the clear asymmetry of some of them (Fig 4 and Table S1), could suggest that the ALH 84001 carbonates are a mixture of several phases, or compositionally heterogeneous on the submicroscopic scale (Cooney et al., 1999). However, our results suggest the presence of a single and compositionally unusual mineral with varying compositions, like the carbonates described by Treiman et al. (2002). This phase has yet to be properly characterized, and doing so will require further work using techniques such as X-ray diffraction and transmission electron microscopy. Raman spectroscopy has demonstrated the presence of hematite and magnetite in those carbonates that were analyzed for the present study (Fig. 4). Hematite affects the positions of several carbonate peaks, and therefore should be taken into account carefully when studying ALH 84001. The presence of magnetite, whose most prominent peak lies at ~660 $cm^{-1}$, is consistent with results of previous studies (McKay et al., 1996; Cooney et al., 1999), although it seems to be present throughout these carbonates (Fig. 4).

As previously noted, our evidence suggests that ALH 84001 carbonates precipitated in at least two different episodes. Consistently, the fracturing and local corrosion of the euhedral crystals of layer #4 prior to precipitation of layer #5 (Fig. 6, D5.3-D5.6) indicate that layers #1 to #4 correspond to the first stage of precipitation, while layers #5 to #7 formed in the second stage and as an overgrowth. Moreover, the sharp transition in carbonate compositions between #4 and #5 suggests a change in solution chemistry and saturation state (e.g., the concentration in $CO_2$ or pH), between formation of the inner (#1 to #4) and outer (#5 to #7) layers. The microprobe data reveal two trends (Figs. 9, 10) that add further weight to the idea of at least two episodes of precipitation. Indeed, the contrasting compositional trends imply that the ambient fluids differed in chemistry, pointing to distinct aqueous alteration events (Fig. 6). The distribution coefficients for Fe and Mn into carbonates are much higher than 1, meaning that any $Fe^{2+}$ or $Mn^{2+}$ in the fluid will preferentially partition into the carbonate (e.g. Chang et al., 1996; Rimstidt et al., 1998). Moreover, the distribution coefficients are not strongly temperature dependent due to the fact that all solubility products of carbonate minerals change with temperature in almost the same way, at least between 273 and 373 K (Rimstidt et al., 1998). Assuming that the system was closed so that the aqueous solutions were not being continually replenished, Fe and Mn would progressively be depleted relative to Mg as carbonate precipitated from this fluid. This process may be responsible for the chemical patterns observed between layers #2 to #4 and layers #5 to #6 (Figs. 9, 10). Layer #1 probably represents an initial precipitation, from a relatively Mg- and Fe-rich fluid. Once the concentration on Fe became lower, the precipitation of a more Ca- and Mn-rich carbonate started (layer #2). After that, successive layers are progressively depleted in Fe and Mn and enriched in Mg (layers #2 to #4, and posteriorly #5 to #6), representing precipitation from an evolving fluid, which was progressively depleted on Fe and Mn as carbonate precipitated. Thus, the sharp change from Fe- and Mn-poor layer #4 to the Fe- and Mn-rich layer #5, would imply the input of a new fluid. Indeed, the presence of small fractures and local corrosion formed prior to layer #5 (Fig. 6) strongly suggest that a new hydrous event introduced a Fe-rich fluid into the system, responsible for the precipitation of layers #5 and #6. The possibility of an additional fluid influx between layers #6 and #7 (layer #7 has higher Fe and Mn content than layer #6) cannot be discounted. This interpretation is in accordance with experimental data obtained by Golden et al. (2001), who performed a multiple step precipitation process to produce chemically and mineralogically zoned synthetic carbonate globules (see their Table 1 and Figure 1). In the first step they added Ca, Fe, and Mg to the starting solution, and as a result a globule with a Ca-Fe rich core precipitated, progressively grading outwards to a more Mg-rich carbonate.

Small high atomic number inclusions occur between carbonate layers #4 and #5 (Fig. 6, D5.5), but their small size prevented identification. We believe that they could be the pyrrhotite and/or magnetite inclusions that other authors have observed in carbonate rosettes rims, which are interpreted to have possibly precipitated when the carbonate was invaded and partially melted by shock-generated liquids (Scott et al., 1997a; McKay and Lofgren, 1997; Bradley et al., 1998). Indeed, the small fractures observed in Fig. 6 (red arrows) are consistent with shock between the precipitation of layers #4 and #5. These apparently amorphous magnetite inclusions, intermixed with iron sulfide, and the voids where they formed, can be the direct result of local carbonate decomposition by an impact (Scott et al., 1997b; Barber and Scott; 2002). However, a strong shock melting of the carbonates would make the borders of layer #4 to be corroded and anhedral, but the boundary between layers #4 and #5 is mostly euhedral. The melting produced by shock should then have been of sub-micrometer scale.

Microprobe data reveal that the studied carbonates fall between dolomite, ankerite, siderite and magnesite (Figs. 5 and 10, and Tables 1 and 2) whereas calcite, the most common terrestrial carbonate, is absent. The compositions of ALH 84001,82 carbonates do not correspond to any known mineral (Harvey and McSween, 1996). Also, there is no solid solution series in terrestrial carbonates from calcite to dolomite or ankerite, and nor from those minerals to magnesite or siderite (see e.g. Chang et al., 1996). Therefore there is no mineral intermediate between these phases, because the substitution of Ca by Mg or Fe is not straightforward. These carbonates with intermediate compositions are unusual on Earth, and their presence suggests a terrestrially uncommon formation scenario. For example, different and variable aqueous solutions under changing environmental circumstances (see for example Fernández-Remolar et al., 2011) supported by kinetic stabilization of these metastable compositions (McKay and Lofgren, 1997).. Compositionally similar carbonate globules have been found on Earth in specific contexts like the volcanic centers in Spitsbergen (Norway), and other few places (Treiman et al., 2002, and references therein). Those carbonates are also described as discs, with cores called 'ASM' due to their compositions lying between ankerite, siderite and magnesite (Treiman et al., 2002).

Our findings suggest that, over time, fluids with different compositions soaked the host rock of this meteorite, possibly indicating distinct aqueous alteration events. This hypothesis should be further explored in the light of new data from orbiters and rovers, and laboratory analysis of meteorites.

## CONCLUSIONS

ALH 84001 is the oldest (4.1 Gyr old) known Martian meteorite and so is a unique source of information on the environmental conditions of early Mars. The main conclusions from this study are as follows:

1) The presence of two sets of fractures in this igneous achondrite support previous studies showing that its constituent minerals have suffered at least two shock events.

2) Carbonate globules, located between cracks in the rock, formed by precipitation from a Mg- and Fe-rich aqueous solution. The euhedral shape observed at the boundary between layers #4 and #5, together with the sharp change in composition, implies precipitation during at least two episodes and from different generations of fluids. Small inclusions at the margins of layer #4, described by other authors as magnetite and sulphide, together with small fractures that do not extend from layer #4 to #5, are explained by a shock event that created the fractures and introduced a second fluid, which precipitated further carbonate as layers #5 and #6, at least. Layer #7 could be the result of a third precipitation event from a later fluid, but the evidence is less clear.

3) The evidence compiled here demonstrates a temporally differentiated fluid environment within the region of the Martian crust that was sampled by ALH 84001. Two, or possibly three, generations of fluids formed the carbonates in different stages. Carbonate precipitation incrementally changed the aqueous environment to promote the precipitation of more Fe- and Mg-rich carbonates. Our model is consistent with previous studies suggesting a hydrothermal origin for the carbonates.

4) ALH 84001 carbonates have been previously related to specific minerals, such as siderite and dolomite. In common with some previous studies, our data suggest the presence of a mineral with composition that is uncommon on Earth. This phase evolves across the immiscibility areas between Fe-, Mg- and Ca-rich carbonates, constituting metastable minerals that require a more comprehensive characterization, opening the door to future analysis on ALH 84001.

5) CL is a very useful technique for locating and characterising carbonates in ALH 84001, and specific compositional zones within them, and it should be considered for use in studies of other carbonate-bearing Martian meteorites.

## ACKNOWLEDGEMENTS

This research has been funded by the Spanish Ministry of Science and Innovation (projects: AYA2011-26522, AYA 2015-67175-P, CTQ2015-62635-ERC and CTQ2014-60119-P to which J.M. Trigo-Rodríguez and C.E. Moyano-Cambero acknowledge financial support). The UK Science and Technology Facilities Council is also thanked for funding through grants ST/H002960/1, ST/K000942/1 and ST/L002167/1. ICN2 and ICMAB acknowledge support of the Spanish MINECO through the Severo Ochoa Centers of Excellence Program under Grants SEV-2013-0295 and SEV-2015-0496, respectively. We acknowledge B. Ballesteros and M. Rosado from the ICN2 Electron Microscopy Division, and to A. Fernández from the ICTS (National Center of Electronic Microscopy) for the SEM, EDS and microprobe measurements. We also thank the NASA Meteorite Working Group, and the Johnson Space Center for providing the ALH 84001,82 section. This study was done in the frame of a PhD. on Physics at the Autonomous University of Barcelona (UAB).

**TABLES**

**Table 1.** Mean and range in chemical composition (maximum and minimum values) for the globule layers.

| Layer | n | MgCO$_3$ | | FeCO$_3$ | | CaCO$_3$ | | MnCO$_3$ | |
|---|---|---|---|---|---|---|---|---|---|
| | | mean | range | mean | range | mean | range | mean | range |
| **1** | 5 | 50.3 | 50.9 - 49.5 | 32.5 | 33.3 - 32.2 | 14.5 | 14.7 - 14.2 | 2.4 | 2.6 - 2.2 |
| **2** | 4 | 40.6 | 42.7 - 38.4 | 29.3 | 29.9 - 27.8 | 24.0 | 25.6 - 22.8 | 5.6 | 7.3 - 3.2 |
| **3.1** | 16 | 48.8 | 54.8 - 45.4 | 33.7 | 35.6 - 30.5 | 14.2 | 16.6 - 12.1 | 2.7 | 4.5 - 2.0 |
| **3.2** | 25 | 61.0 | 71.6 - 50.2 | 28.7 | 34.0 - 20.5 | 9.1 | 13.2 - 2.7 | 0.7 | 1.9 - 0.2 |
| **3.3** | 7 | 77.1 | 85.5 - 70.6 | 16.9 | 22.6 - 8.8 | 5.0 | 6.3 - 3.0 | 0.4 | 0.9 - 0.2 |
| **4** | 12 | 81.7 | 89.1 - 64.0 | 12.5 | 27.3 - 6.8 | 4.7 | 7.4 - 2.7 | 0.3 | 0.6 - 0.1 |
| **5** | 19 | 49.8 | 67.5 - 35.0 | 40.1 | 51.9 - 27.4 | 7.9 | 10.7 - 2.0 | 1.2 | 2 - 0.2 |
| **6** | 21 | 90.9 | 95.1 - 78.3 | 4.4 | 14.7 - 0.3 | 4.3 | 6.5 - 3.2 | 0.2 | 0.6 - 0.0 |
| **7** | 7 | 76.5 | 84.7 - 68.8 | 16.0 | 23.1 - 9.2 | 6.3 | 6.8 - 5.5 | 0.5 | 0.8 - 0.4 |

Data obtained by electron microprobe.
The values are mole % carbonates.
n = number of analyses.

**Table 2.** Chemical compositions obtained from the carbonate globule analyses in Figure 5.

| Point | Layer | MgCO$_3$ | FeCO$_3$ | CaCO$_3$ | MnCO$_3$ |
|---|---|---|---|---|---|
| **31** (J7.2) | #3 | 57.91 | 30.89 | 10.38 | 0.81 |
| **34** (J7.2) | #3 | 57.86 | 31.04 | 10.33 | 0.77 |
| **38** (J7.2) | #3 | 71.00 | 22.19 | 6.38 | 0.43 |
| **12** (J7.2) | #5 | 36.21 | 52.07 | 9.97 | 1.75 |
| **49** (J7.2) | #6 | 89.26 | 7.46 | 3.26 | 0.01 |
| **13** (J7.2) | #6 | 91.80 | 4.75 | 3.41 | 0.04 |
| **53** (J7.2) | #7 | 81.94 | 11.62 | 6.01 | 0.43 |
| **3** (J7.3) | #1 | 50.80 | 32.43 | 14.53 | 2.23 |
| **4** (J7.3) | #1 | 50.41 | 32.38 | 14.60 | 2.60 |
| **6** (J7.3) | #2 | 45.18 | 33.68 | 16.63 | 4.52 |
| **27** (J7.3) | #2 | 40.31 | 30.15 | 22.96 | 6.58 |
| **5** (J7.3) | #2 | 38.21 | 29.86 | 24.64 | 7.29 |
| **7** (J7.3) | #3 | 50.20 | 32.25 | 15.01 | 2.53 |
| **82** (D5.1) | #3 | 60.86 | 30.02 | 8.87 | 0.25 |
| **84** (D5.1) | #3 | 66.24 | 25.95 | 7.26 | 0.54 |

| 87 (D5.1) | #4 | 87.79 | 8.39 | 3.46 | 0.35 |
| 92 (D5.1) | #5 | 60.49 | 32.45 | 6.17 | 0.89 |
| 97 (D5.1) | #6 | 86.65 | 8.62 | 4.62 | 0.11 |
| 98 (D5.1) | #6 | 93.96 | 1.11 | 4.79 | 0.13 |
| 101 (D5.1) | #7 | 77.37 | 15.58 | 6.27 | 0.77 |
| 118 (D5.2) | #3 | 65.34 | 27.21 | 7.05 | 0.40 |
| 119 (D5.2) | #4 | 85.06 | 9.61 | 5.19 | 0.14 |
| 123 (D5.2) | #5 | 43.46 | 45.43 | 9.31 | 1.80 |
| 126 (D5.2) | #6 | 91.49 | 4.86 | 3.56 | 0.09 |
| 128 (D5.2) | #6 | 92.51 | 0.78 | 6.52 | 0.19 |

Data obtained by electron microprobe.
The values are mole % carbonates.

**FIGURE CAPTIONS**

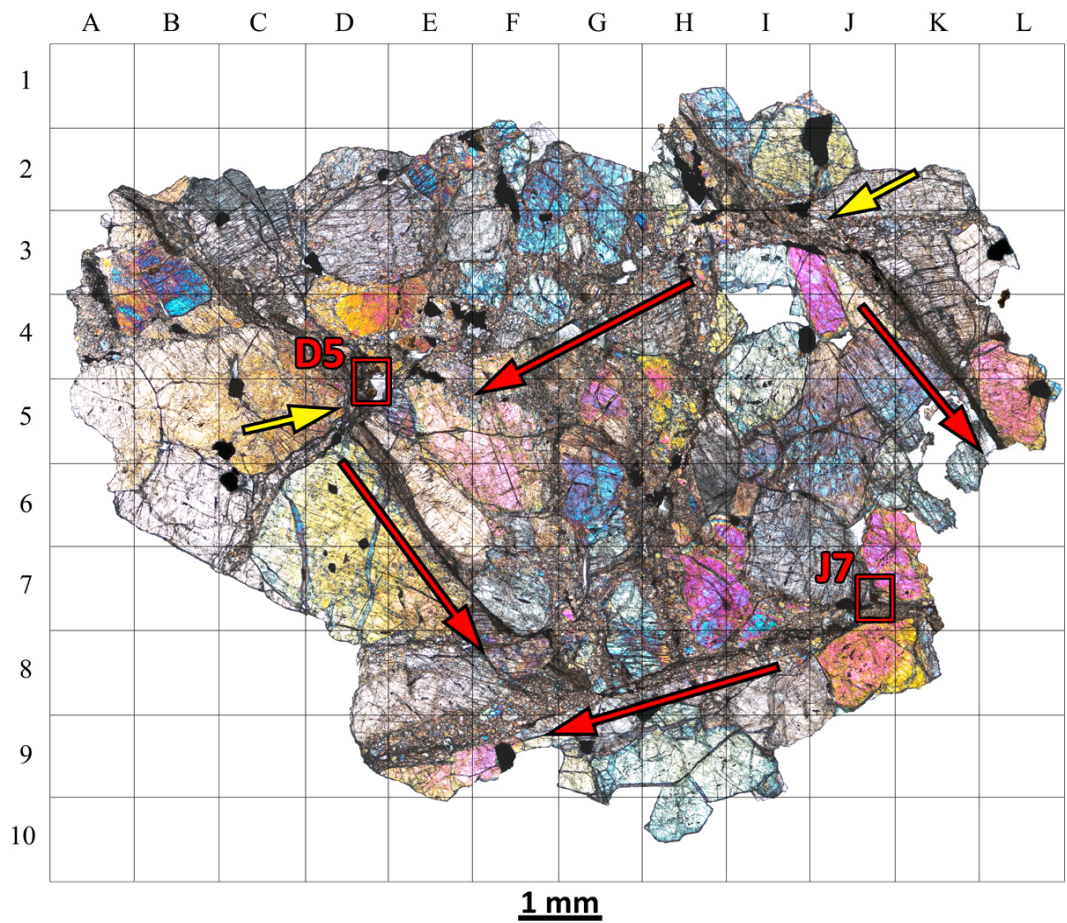

1 mm

**Fig. 1.** High resolution mosaic of the ALH 84001,82 thin section (images acquired in TL between crossed polarisers). Red arrows show the orientations of the two main fracture systems. Yellow arrows point to where the fractures intersect. Fractures oriented NNW-SSE are displaced by those aligned ENE-WSW. D5 and J7 are the two carbonate-containing regions studied.

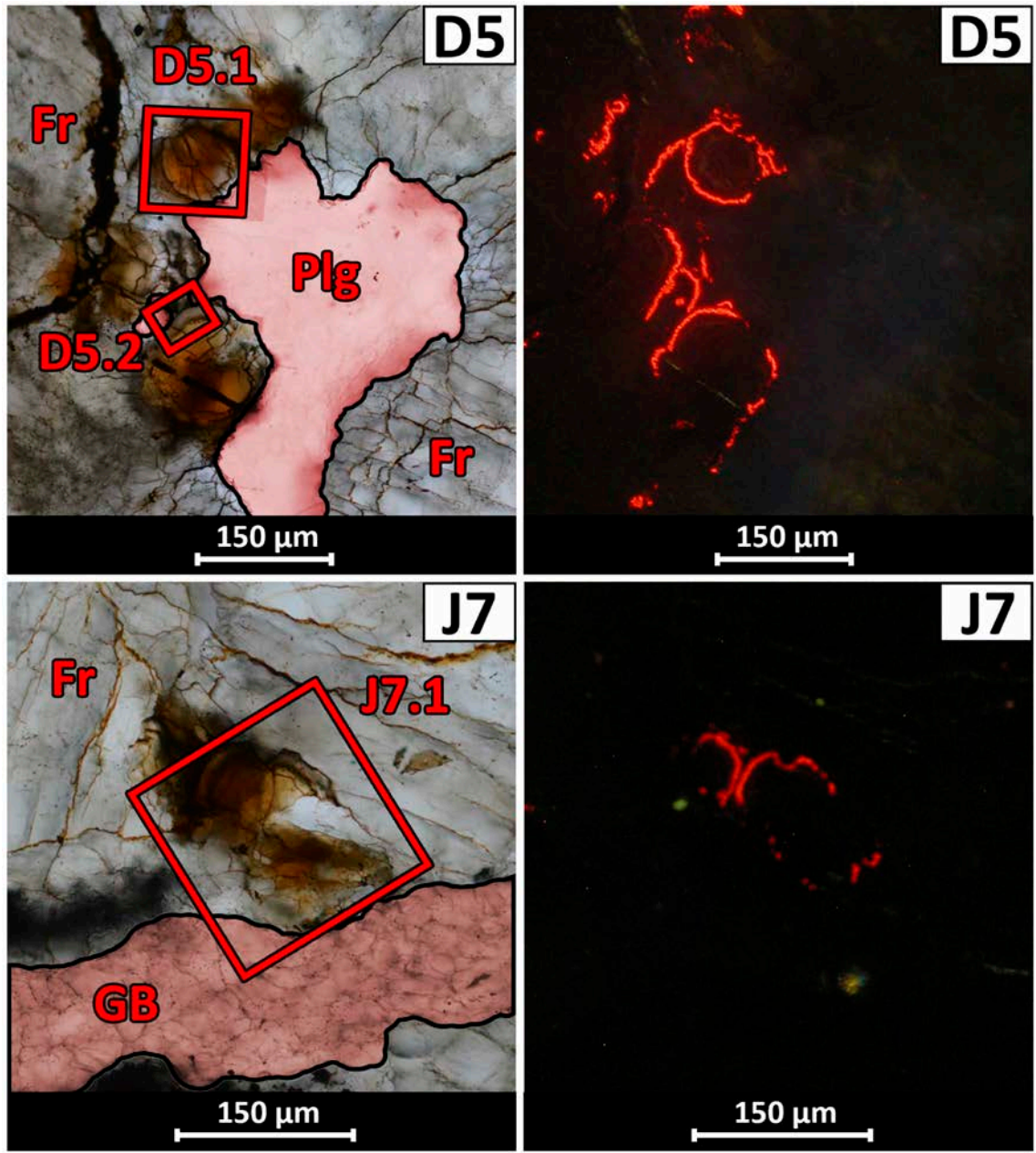

**Fig. 2.** Detail of the two regions of interest highlighted in Figures 1 and S1 (D5 and J7); TL left, CL right. Carbonates are the brown to orange areas in TL, while their Fe-poor and Mn-rich layers are the red rims in CL. In the TL images we draw attention to the fractures around regions where the carbonates are present (Fr). The maskelynite or shock modified plagioclase glass solidified in a fracture or pore after formation of the carbonates at D5 (Plg, highlighted in red), and in one of the so-called granular bands (Mittlefehldt, 1997) next to the carbonate at J7 (GB, painted in red). Three areas that are in some of the later Figures (D5.1 and J7.1) are also highlighted.

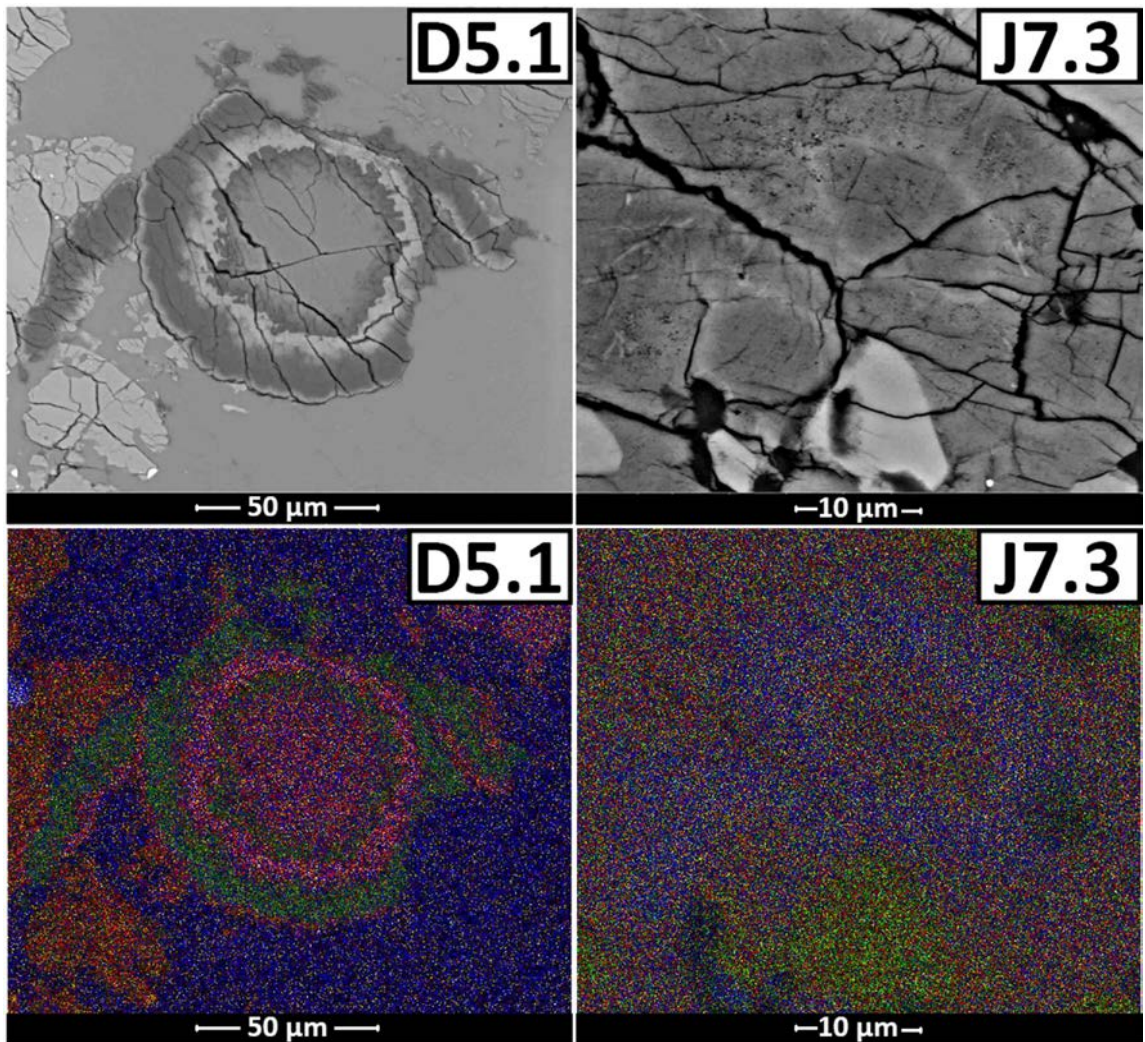

**Fig. 3.** BSE images of the areas D5.1 and J7.3 (see Figs. 2 and 4 for the location of J7.3) with corresponding EDS maps revealing the distribution of Ca (blue), Fe (red), Mg (green) and Mn (yellow). In D5.1, the Fe-rich and Mg-rich rims of the carbonate can be clearly observed. Also note that the outer layers of the globules are partially corroded, and that the surrounding mineral (maskelynite, or shock modified plagioclase) incorporates fragments of the carbonate. In J7.3 the carbonate is compositionally more homogeneous, apart for the most outer part of layer #2 (see J7.3 in Fig. 7), whose Ca-rich composition is shown more clearly by the the EDS map than by the BSE image.

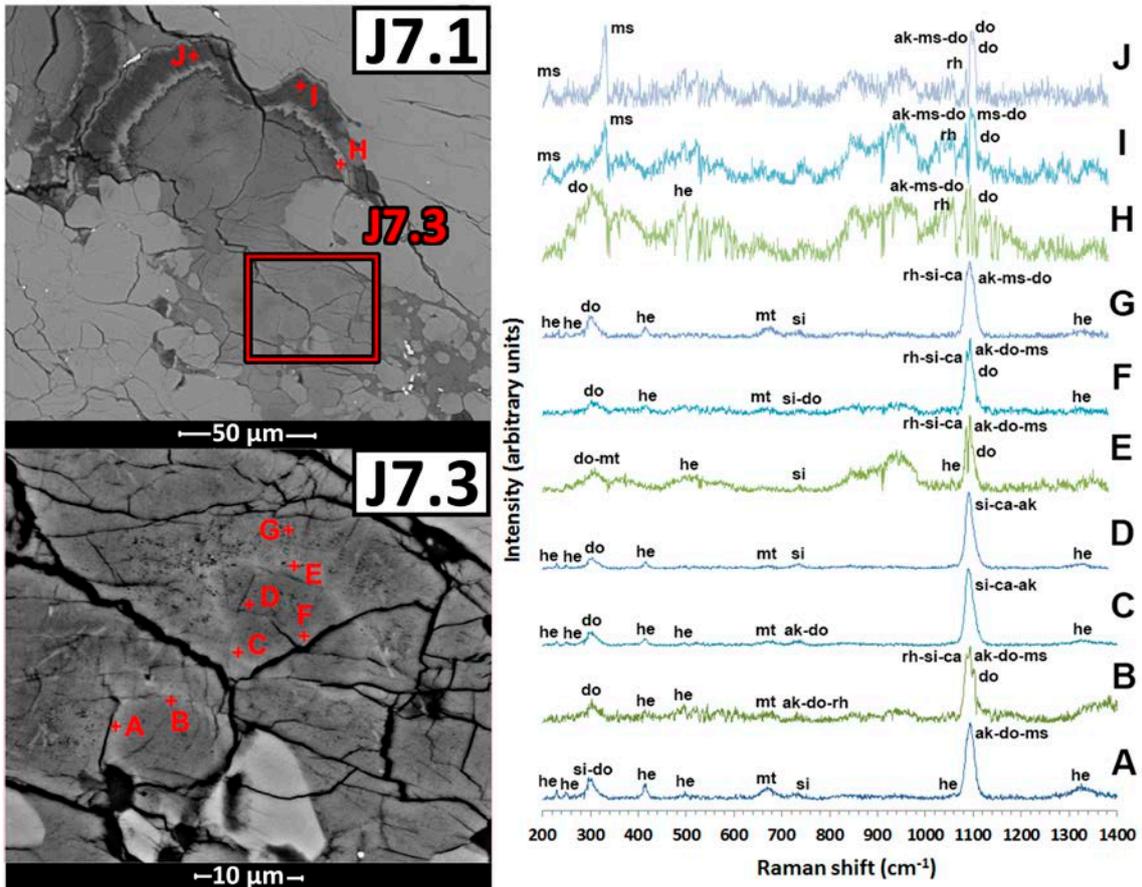

**Fig. 4.** BSE images of areas J7.1 and J7.3 (see Fig. 2), where Raman spectroscopy was performed (left) and corresponding Raman spectra (right). The minerals that match each peak have been suggested (according to the data in the RRUFF catalogue, Downs, 2006): ms (magnesite), ak (ankerite), do (dolomite), rh (rhodochrosite), he (hematite), mt (magnetite), si (siderite), ca (calcite). The positions of the peaks are listed in Table S1.

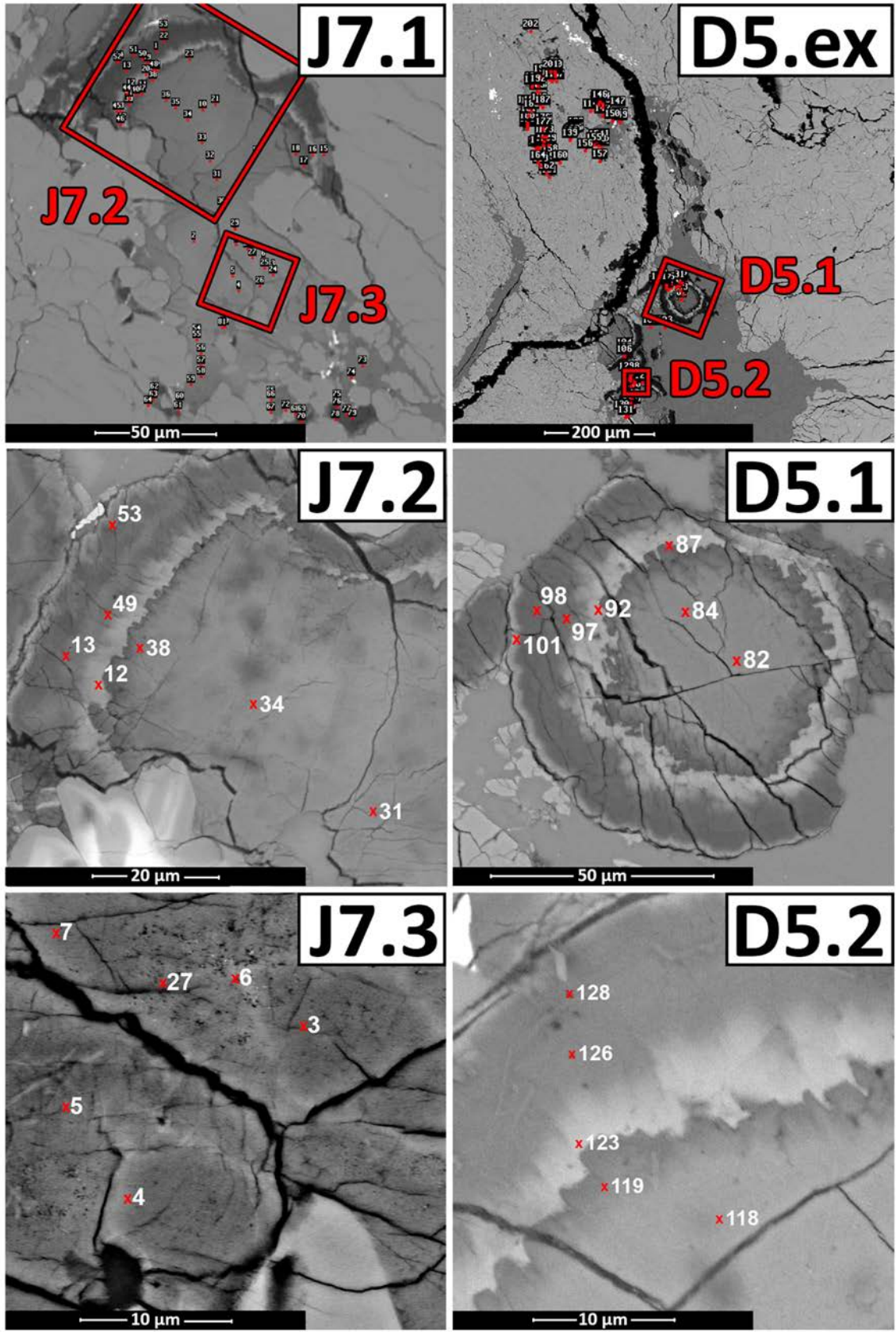

**Fig. 5.** BSE images of the carbonate globules studied, including the locations of the 160 microprobe spots. J7.1 is highlighted in Fig. 2, and J7.2 and J7.3 are showing detail of this area. D5.ex is a region showing D5 (see Figs. 1, 2 and S1) but also the area above, where some carbonates were analysed. D5.1 and D5.2 are also showing specific areas in greater detail, and both were already highlighted in Fig. 2. All the data obtained are shown in Fig. 10. Only some example points can be seen in J7.2, J7.3, D5.1, and D5.2; the data obtained from these points are listed in Table 2.

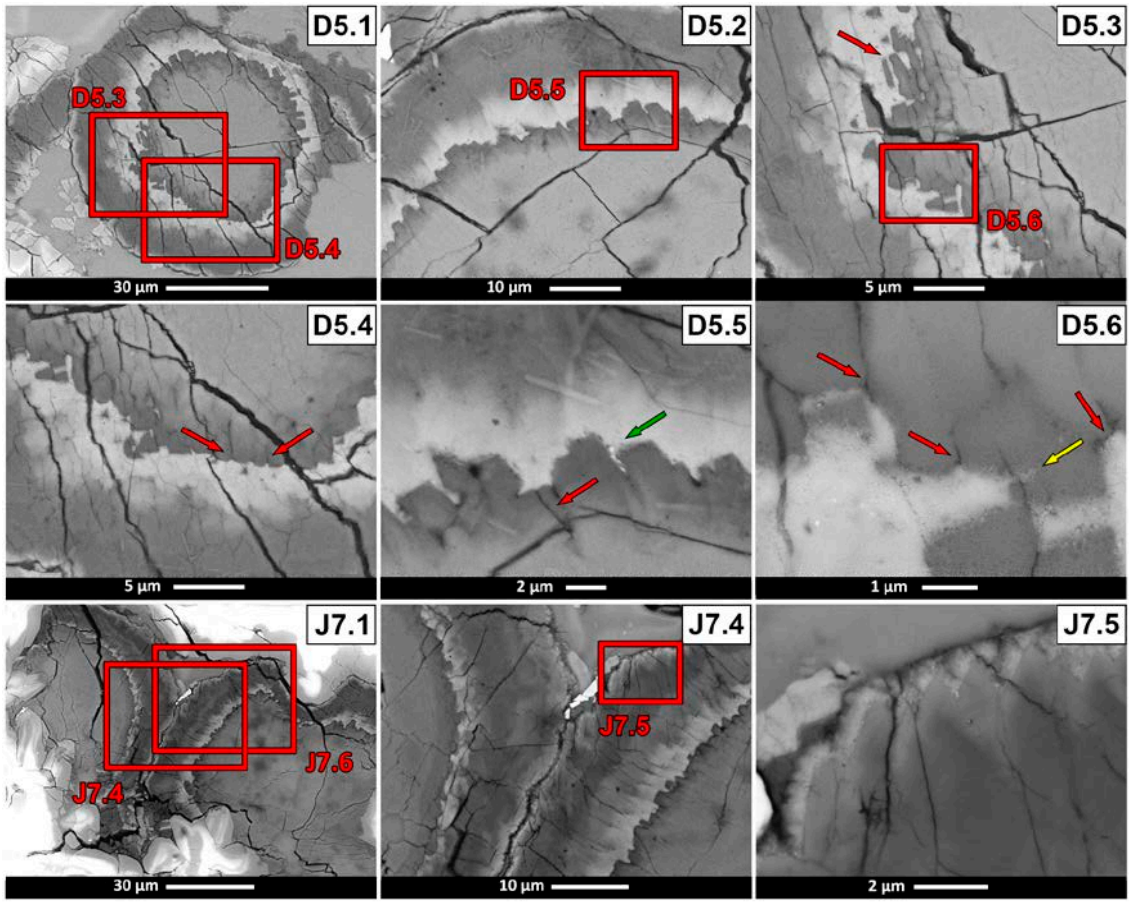

**Fig. 6.** Detailed BSE images showing the boundary features between layers #4 and #5 (D5.1-D5.6) on carbonate globules (see also D5.1 in Fig. 7). Red arrows point to the small fractures that terminate at the end of layer #4. The green arrow indicates small high atomic number inclusions. The yellow arrow shows corrosion at the end of layer #4. J7.1, J7.4 and J7.5 show the complex gradation in colour (and composition) between layers #5 and #7 (see J7.6 in Fig. 7).

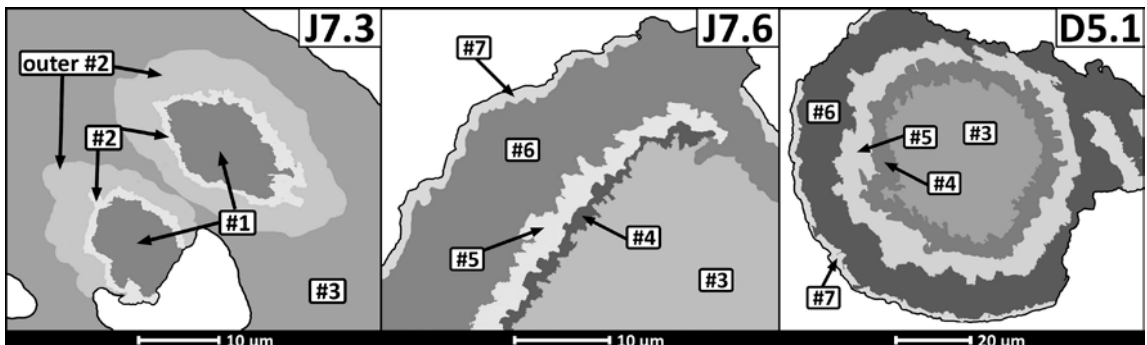

**Fig. 7.** Schematic images showing the different layers in the carbonates studied. The area indicated as 'outer #2' is the portion of layer #2 that is distinguishable as a blue ring in the element map (Fig. 3).

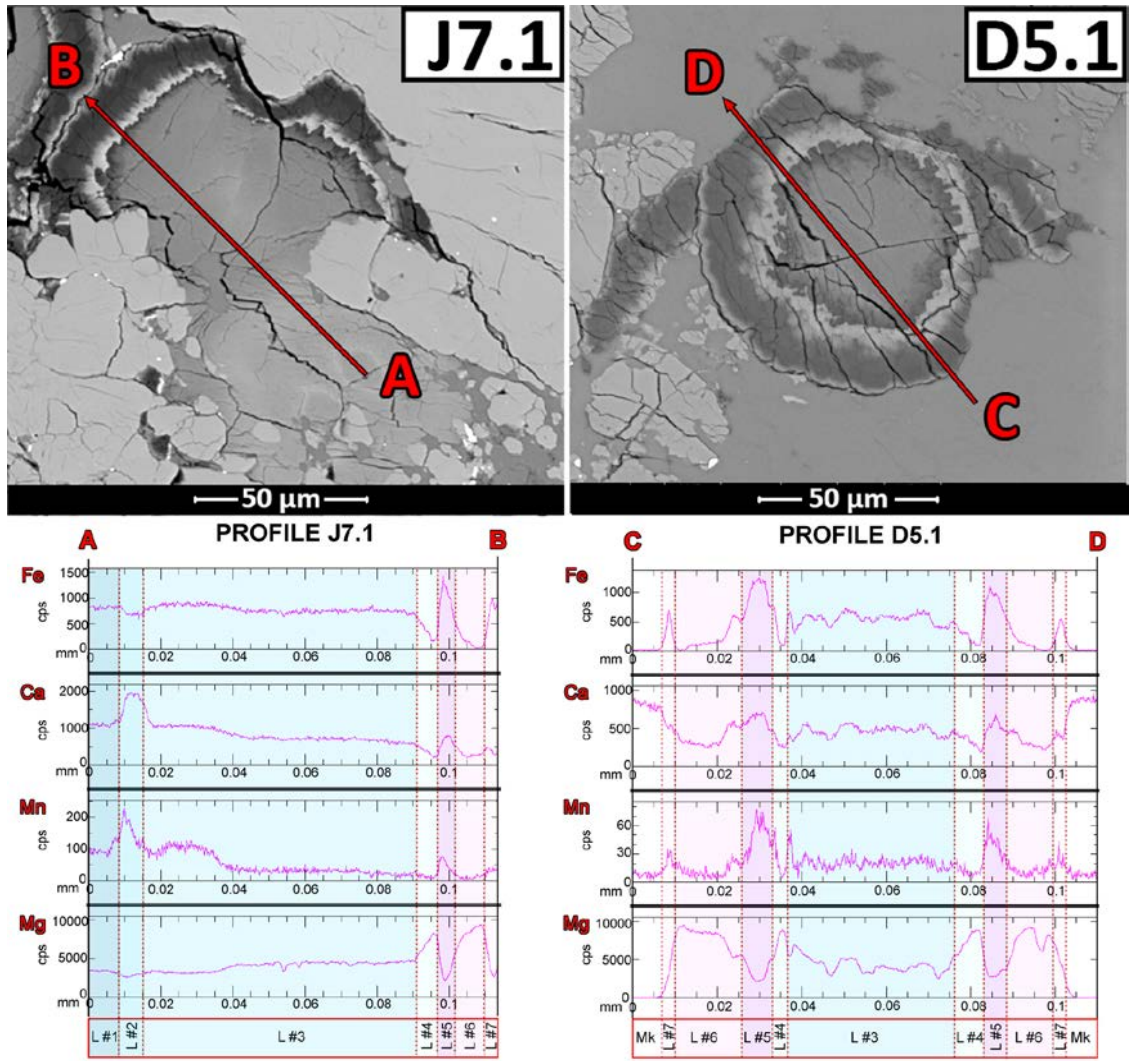

**Fig. 8**. Elemental profiles obtained by electron microprobe showing progressive geochemical variations along two globules. In the upper part the lines along which the profiles where obtained are indicated, in J7.1 and D5.1. Cps: counts per second; L: layer; Mk: maskelynite (or shock modified plagioclase glass).

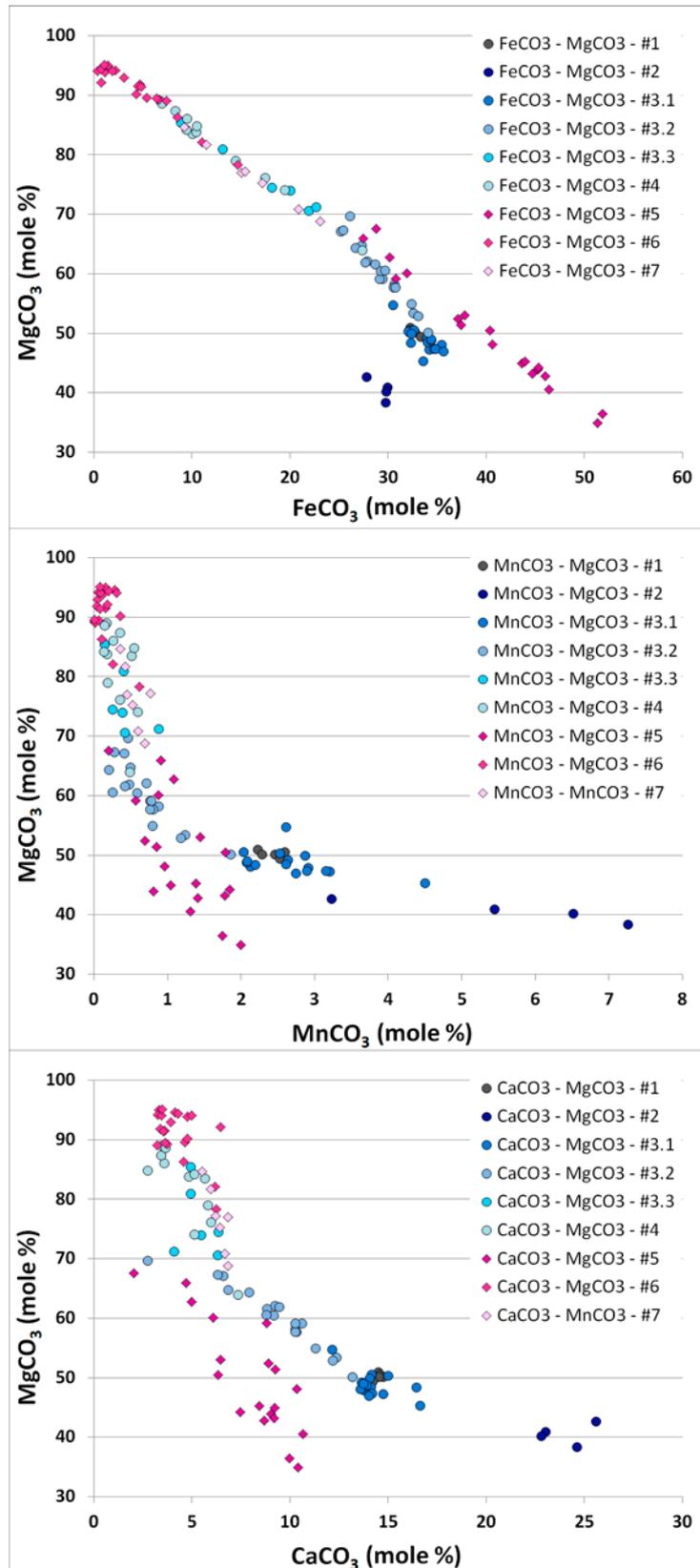

**Fig. 9**. Plots of MgCO₃ versus FeCO₃, MnCO₃ and CaCO₃ (in mole %) derived from microprobe analyses (see J7.1 and D5.ex in Fig. 5). Each layer is marked with a different colour. Also, layers #1 and #4 (before the euhedral shape), are represented by bluish circles, while layers #5 to #7 are represented by purple to pink diamonds. As compositions vary within layer #3 (see Fig. 7), it is subdivided into #3.1 (slightly

clearer, just after layer #2), #3.2 (intermediate between layers #2 and #4) and #3.3 (slightly darker, just before layer #4).

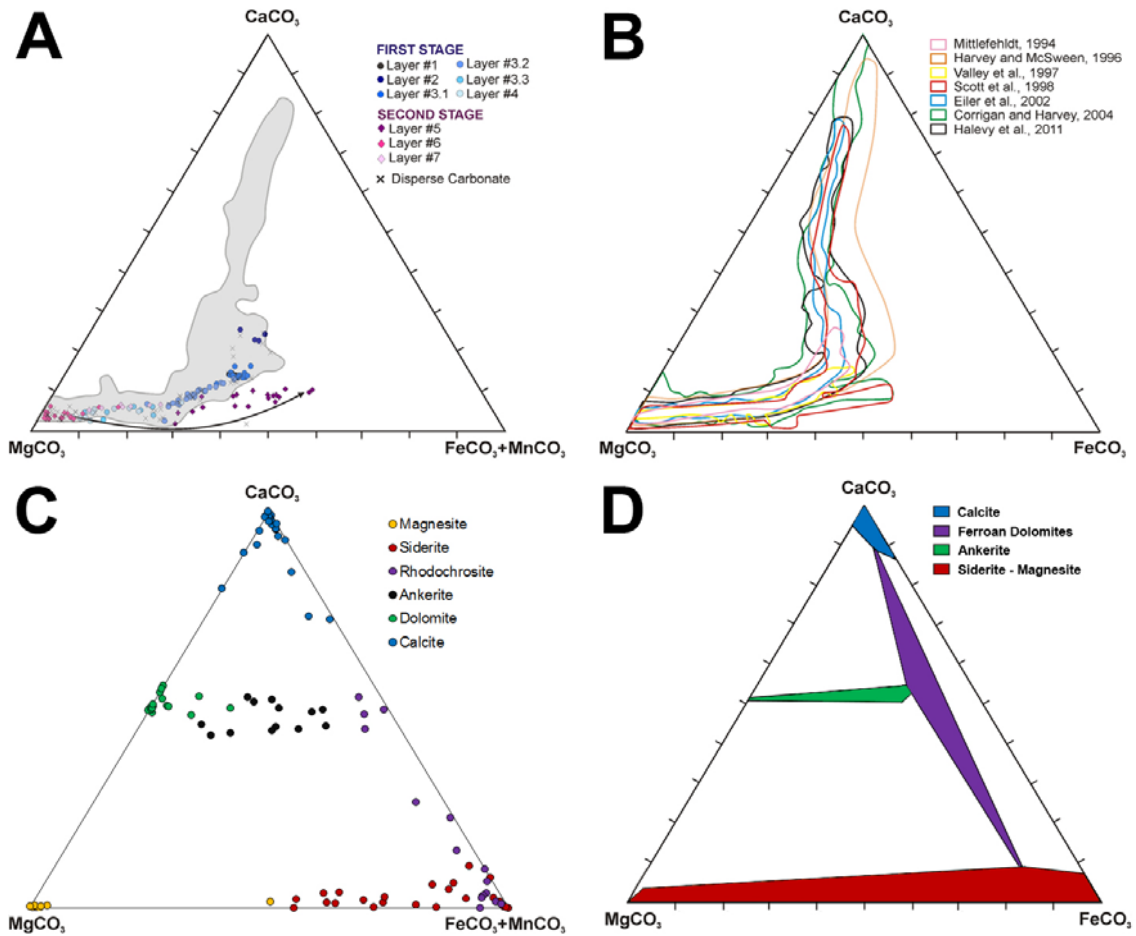

**Fig. 10. A.** Ternary diagrams showing the data obtained by electron microprobe from around 160 different spots (see J7.1 and D5.ex in Fig. 5 and Table 1). Note the progressive compositional change between layers #1 and #4 and the sharp compositional change between layers #4 and #5 (black arrow). The shaded are shows published carbonate compositions of ALH 84001 (based on data published by Holland et al., 1999; 2005, who plotted Fe+Mn contents). **B.** Ternary diagram showing, in different colors, carbonate compositions of ALH 84001 published by different authors (Mittlefehldt, 1994; Harvey and McSween, 1996; Valley et al., 1997; Scott, 1998; Eiler et al., 2002; Corrigan and Harvey, 2004; Halevy et al., 2011), who only plotted Fe contents. **C.** Compositions of several terrestrial carbonates (data taken from Chang et al., 1996). **D.** Distribution of the calcite solid solutions (blue), ferroan dolomites with less than 20% Fe substitution (purple), ankerite solid solutions (green), and siderite-magnesite solid solutions (red), at 450ºC (adapted from Rosenberg, 1967). The diagrams show the mole % of $MgCO_3$, $CaCO_3$, $FeCO_3$ and $MnCO_3$.

### SUPPLEMENT

Supporting material for "Petrographic and geochemical evidence for multiphase formation of carbonates in the Martian orthopyroxenite Allan Hills 84001" by Moyano-Cambero et al.

# SUPPLEMENTARY FIGURE

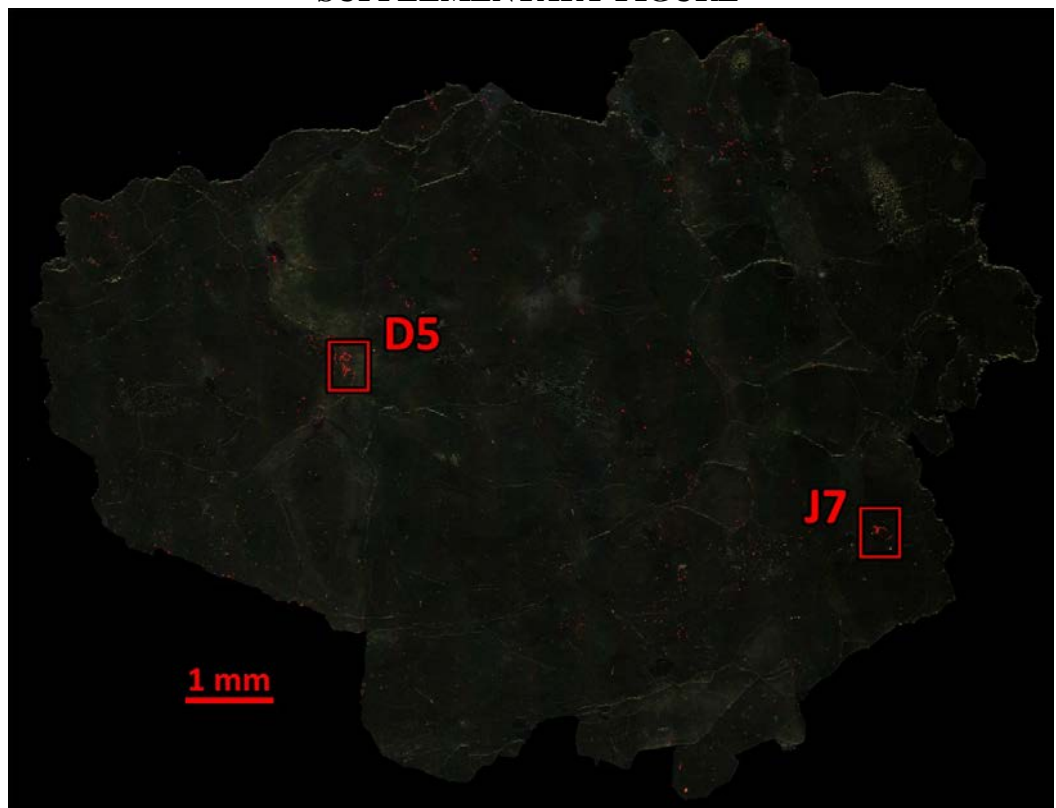

**Fig. S1.** High resolution CL mosaic of the ALH 84001,82 thin section that can be compared directly with Figure 1. Bright red areas correspond to very Fe-poor and Mn-rich layers in carbonates.

# SUPPLEMENTARY TABLES

**Table SI.** Position (in cm$^{-1}$) of the main peaks found by Raman spectroscopy (see Fig. 5). Those in bold were clearly distinguishable from noise before the background was removed. Other minor peaks, plus some whose origin has not been determined, are not listed.

| Point | Detected Peaks on Each Point | | | | | | | | | | | | | |
|---|---|---|---|---|---|---|---|---|---|---|---|---|---|---|
| A | n.d. | **230** | **250** | **296** | **415** | **498** | **670** | **733** | n.d. | n.d. | **1094** | n.d. | n.d. | **1326** |
| B | n.d. | 229 | 253 | **303** | **414** | **496** | **679** | **722** | n.d. | **1086** | **1094** | **1100** | **1103** | n.d. |
| C | n.d. | **229** | **250** | **303** | **413** | **497** | **668** | **725** | n.d. | n.d. | **1090** | n.d. | n.d. | **1326** |
| D | n.d. | **229** | **249** | **303** | **416** | 500 | **681** | **735** | n.d. | n.d. | **1090** | n.d. | n.d. | **1323** |
| E | n.d. | n.d. | 247 | **310** | n.d. | **507** | 658 | **734** | **1056** | **1086** | **1093** | **1099** | n.d. | n.d. |
| F | n.d. | n.d. | n.d. | **303** | **415** | 497 | **670** | **730** | n.d. | **1087** | **1093** | n.d. | **1102** | **1314** |
| G | n.d. | **234** | **249** | **302** | **415** | n.d. | **675** | **734** | n.d. | **1088** | **1093** | n.d. | n.d. | **1333** |
| H | n.d. | n.d. | n.d. | **303** | n.d. | **497** | 658 | 734 | 1057 | **1085** | **1093** | 1094 | **1103** | 1332 |
| I | **216** | n.d. | n.d. | **331** | n.d. | 497 | 660 | n.d. | 1056 | **1085** | **1095** | **1098** | **1103** | 1330 |
| J | **214** | n.d. | 252 | **332** | n.d. | 497 | 663 | 736 | 1055 | **1085** | **1096** | **1100** | **1102** | 1326 |